\documentclass[]{spie}  
\usepackage[]{graphicx}
\usepackage{siunitx}
\usepackage{subcaption}	
\usepackage{sidecap}
\usepackage{wrapfig}

\DeclareSIUnit{\cmps}{\cm\per\second}
\DeclareSIUnit{\mps}{\meter\per\second}
\DeclareSIUnit{\kmps}{\kilo\meter\per\second}
\DeclareSIUnit{\micrometer}{$\mu$\meter}
\DeclareSIUnit{\foot}{'}
\DeclareSIUnit{\inch}{"}

\title{MAROON-X: A Radial Velocity Spectrograph for the Gemini Observatory} 

\author{Andreas Seifahrt\supit{a}, Julian St\"urmer\supit{a}, Jacob L. Bean\supit{a}, Christian Schwab\supit{b,c} 
\skiplinehalf
\supit{a}University of Chicago, USA \\
\supit{b}Macquarie University, Sydney, Australia\\
\supit{c}Australian Astronomical Observatory, Sydney, Australia\\
}

\authorinfo{Further author information: (Send correspondence to A.S.)\\A.S.: E-mail: seifahrt@uchicago.edu, Telephone: 1 773 702 9877}

\begin{document} 
  \maketitle 

\begin{abstract}
MAROON-X is a red-optical, high precision radial velocity spectrograph currently nearing completion and undergoing extensive performance testing at the University of Chicago. The instrument is scheduled to be installed at Gemini North in the first quarter of 2019. MAROON-X will be the only RV spectrograph on a large telescope with full access by the entire US community. In these proceedings we discuss the latest addition of the red wavelength arm and the two science grade detector systems, as well as the design and construction of the telescope front end. We also present results from ongoing RV stability tests in the lab. First results indicate that MAROON-X can be calibrated at the sub-m\,s$^{-1}$ level, and perhaps even much better than that using a simultaneous reference approach.
\end{abstract}


\keywords{Gemini, radial velocity, echelle spectrograph, optical fibers, CCD, active fiber guiding, radial velocity stability, Fabry Perot etalon}
\section{INTRODUCTION}
\label{sec:intro}  
Radial velocities (RVs) dominated the first fifteen years of exoplanet science by delivering the first statistical constraints on planet occurrence around a wide range of stars. With the field of exoplanets now in its third decade, the RV technique remains at the forefront due to its continued unmatched sensitivity to planets around nearby stars and its important supporting role for other exoplanet detection techniques.

In terms of support for other techniques, RV follow-up measurements in support of ground- and space-based transit surveys are a critical component of the confirmation of candidate exoplanets and for their mass measurements. With the successful \textit{TESS} launch this past April and the upcoming launch of the European \textit{CHEOPS} mission by the end of this year (2018), many new small transiting planets will be identified over the next decade. This presents an enormous opportunity to expand our study of planetary statistics into the regime of planet bulk compositions, with the RV method delivering the mass measurements of these objects. These mass measurements also become essential to select suitable targets for follow-up atmospheric studies using \textit{HST}, \textit{Spitzer}, and \textit{JWST} and future direct imaging missions.

The RV technique is also a critical component of the near-term opportunity to study potentially habitable planets orbiting M dwarfs, and in particular the very lowest-mass M dwarfs (those with M$_{\star}$\,$<$\,0.3\,M$_{\odot}$). In contrast to solar-type stars, the habitable zones of these stars are close-in enough so that planets in this region have a significant chance of transiting and would be feasible targets for transit spectroscopy observations to characterize their atmospheres. Although M dwarfs comprise the bulk of the stellar population in the solar neighborhood, they are intrinsically faint. Precision RV measurements of mid to late M dwarfs thus require a large size telescope and an efficient spectrograph.  


With the recent installation of ESPRESSO at the VLT \cite{espresso2018}, HPF at the HET \cite{hpf2018}, and IRD at Subaru\cite{ird2018}, a new generation of fiber-fed, highly stabilized RV spectrographs has arrived at 8-10\,-m class telescopes. In the coming years, more instruments at large telescopes will become available, such KPF on Keck\cite{KPF2018} and iLocator on the LBT\cite{ilocater2016}. However, none of these instruments offers unrestricted access to the entire US community. 

\section{MAROON-X RV spectrograph: Overview}

Over the last few years we have been developing an instrument called MAROON-X to provide the US community with a state-of-the-art RV spectrograph on a telescope large enough to facilitate efficient RV measurements even on very late-type M dwarfs. MAROON-X was originally designed for deployment at the 6.5\,m Magellan telescope at Las Campanas Observatory in Chile, but is now slated to be installed at the 8\,m Gemini North telescope on Mauna Kea, Hawaii. MAROON-X will initially be operated as a visitor instrument with the ultimate goal being full integration as a facility class instrument. The move from Magellan to Gemini increases the instrument's grasp and secures the US community with a RV spectrograph at a large telescope with an open-access policy driven by scientific merit rather than institutional affiliation. 

MAROON-X is a fiber-fed, highly stabilized, high-efficiency spectrograph with a resolving power of $R$=80,000, covering a bandpass of 500 -- 900\,nm in two camera arms. A pupil slicer and double scrambler as well as simultaneous wavelength calibration with a Rb-traced white light etalon\cite{etalon} will allow RV measurements with an intrinsic precision of under 1\,ms$^{-1}$. An updated list of the main characteristics of MAROON-X is given in Table\,\ref{table} and a detailed description of the instrument design can be found in a previous SPIE proceedings paper\cite{maroon-x}. 

Because it was initially designed for a smaller telescope, MAROON-X's 100\,$\mu$m octagonal science fiber will have a FOV of  0.77 arcsec at Gemini, only slightly larger than the 70\%-tile seeing limited FWHM of 0.75 arcsec on Gemini North. The geometric fiber coupling efficiency is thus about 52\%. However, the spectrograph itself has exquisite efficiency of up to 60\% in the blue arm as measured from the fiber exit to the focal plane\cite[Fig. 7]{maroon-x}, and we expect a similar number for the red arm (see Section \ref{redarm}). A new 4k$\times$4k deep-depletion 100\,$\mu$m thick CCD with QE of over 90\% guarantees low fringing and high efficiency (see section \ref{CCDs}) out to 900\,nm. We have designed the fiber-injection unit with a nominal throughput of 86\% and a diffraction-limited image quality on-axis (see section \ref{frontend}). Losses at the pupil slicer and doubler scrambler and due to fiber FRD are estimated to be about 25\%. This leaves a total peak efficiency of about 18\%, which is very competitive among similar instruments. Gemini's silver coated mirrors and small central obstruction will further contribute to the overall efficiency of RV measurements with the instrument. 

\begin{SCfigure}[][!t]
\centering
\vspace{0mm}
\includegraphics[trim={0cm 0cm 12cm 10cm},width=0.6\linewidth,clip]{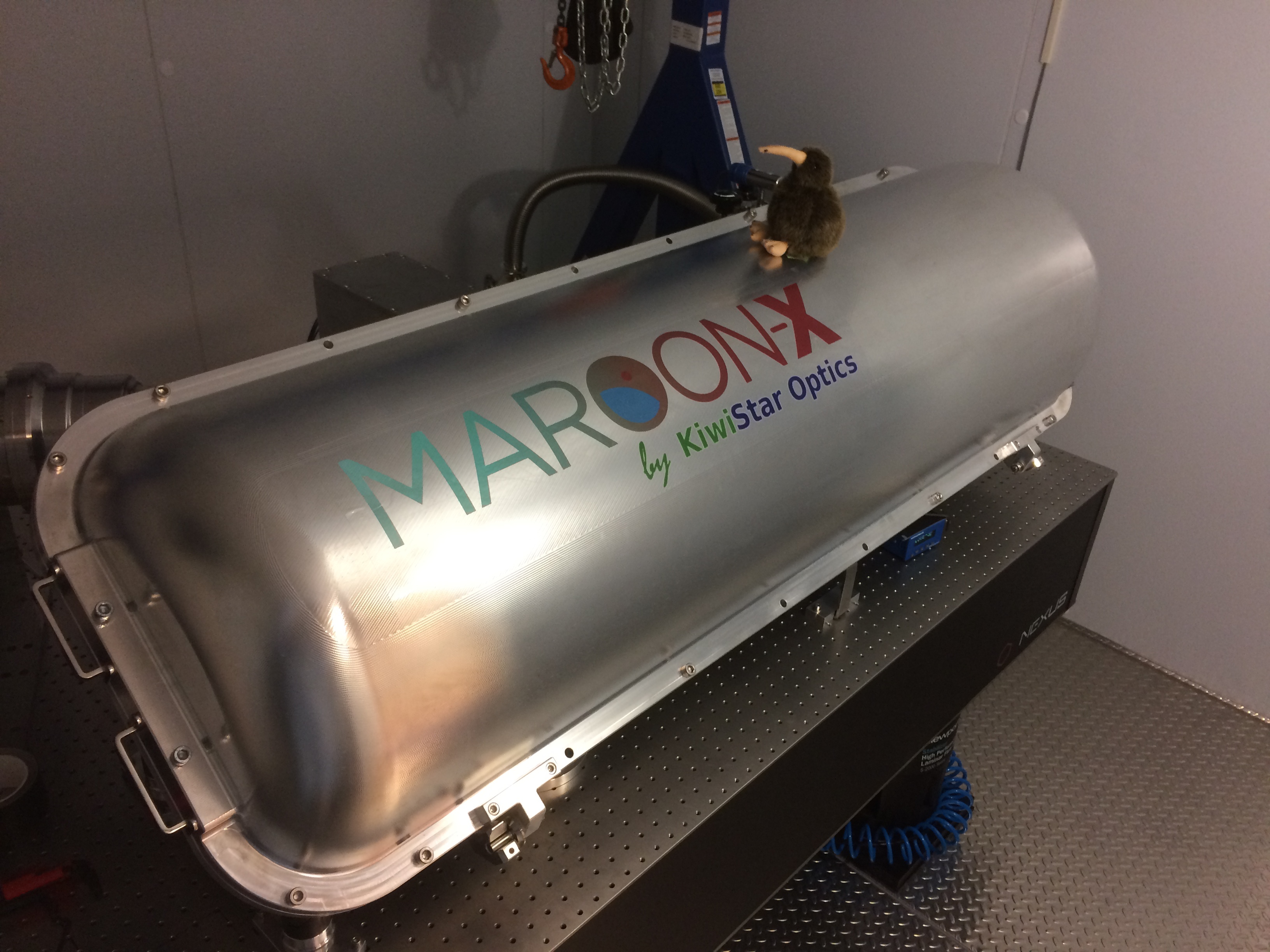}
\vspace{0mm}
\caption{\textbf{Core spectrograph of MAROON-X installed in its environmental chamber in the lab at the
University of Chicago.} The vacuum vessel houses the main spectrograph components while the cross-dispersers and camera arms are in pressure sealed barrels outside the chamber (behind the vacuum vessel, not visible in the image).}
\label{enclosure}
\end{SCfigure}

\section{Red arm}\label{redarm}

Since the funding environment for a project of this size is challenging, we originally ordered the core echelle spectrograph, a variant of the ``KiwiSpec R4-100'' \cite{barnes}, from KiwiStar Optics (New Zealand) with only the blue arm (covering 500--670\,nm) implemented. After taking delivery of the instrument in January 2017, we ordered the missing red arm in the fall of 2017. The red arm will cover the wavelength range of 650--900\,nm. 

The camera optics in the red arm posed some challenges, as the high anamorphic pupil compression of 1:3 hinders an effective aberration compensation which is compounded by the longer orders (and, hence, wider angles) in the red camera compared to the blue camera. Consequently, even after a lengthy optimization procedure, the average spot sizes for the red camera are slightly larger than for the blue camera and the width of  the 80\% EE in dispersion direction ranges from 4\,$\mu$m to almost 20\,$\mu$m compared to an average slit width of 53\,$\mu$m. 

\begin{figure}[!t]
\centering
\vspace{0mm}
\includegraphics[trim={0cm 5cm 0cm 5cm},width=0.75\linewidth,clip]{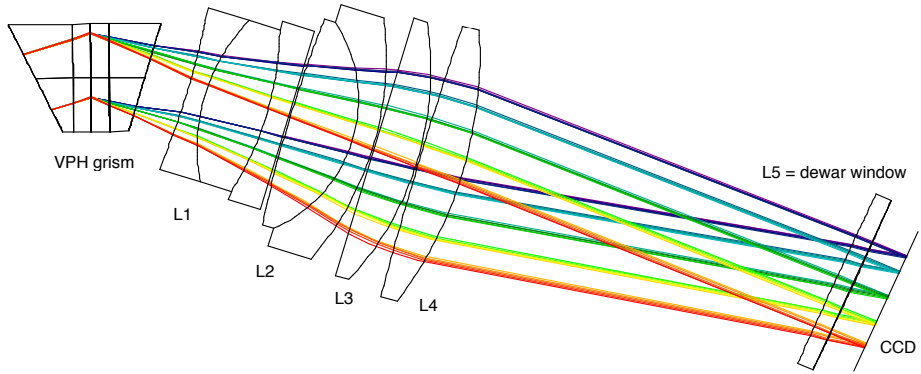}
\vspace{4mm}
\caption{\textbf{Optical design of the red camera and cross-disperser arm for MAROON-X}. A dichroic and fold mirror (not shown here) steer the white pupil onto a VPH grism for cross-dispersion. The camera is comprised of a triplet, a doublet and three singlet lenses. The last lens, L5, is made of fused silica and doubles as the dewar entrance window for the CCD cryostat. The camera has a long back-focal length of over 200mm.}
\label{redcamera}
\end{figure}

The camera design by Damien Johnes (Prime Optics, Australia) uses a combination of five different Ohara glasses as well as fused silica (see Figure \ref{redcamera}) and features a long back-focal length to ease the placement of the CCD cryostat next to the camera and cryostat of the blue arm. The last lens of the red camera (L5) is made of fused silica and has a cylindrical surface facing the camera and a spherical surface facing the CCD detector. It doubles as the dewar entrance window for the CCD cryostat. All lenses have BBAR coatings with an average reflectivity of R$\le$0.5\%.

The cross-disperser is a VPH grism with similar characteristics as the one used in the blue arm. The VPH gratings are made by Kayser Optical Systems Inc. (KOSI). The blue VPH has 1,652\,l\,mm$^{-1}$, while the red VPH has 1,161\,l\,mm$^{-1}$.

The camera and cross-disperser optics are currently in production and we expect delivery of the red arm by September 2018. Integration with the spectrograph will be uncritical as the dichroic and red fold mirror are already installed and the vacuum chamber of the spectrograph does not need be opened for the installation of the new optics.

\section{Science grade detector systems}\label{CCDs}

For the same reasons as the delayed order of the red arm, we had to delay the order of our detector systems until funding became available. In March 2017 we placed an order with STA Inc. to deliver two complete 4k$\times$4k detector systems based on their new STA4850 CCD. The detector systems include cryostats, Bonn shutters, and two water cooled Archon CCD controllers. We expect the delivery of the complete detector systems by August 2018.

The STA4850 is a 4k$\times$4k device with 15\,$\mu$m pixel size. It was designed specifically for MAROON-X as an improvement over the STA4150. It adds a dump drain and back bias features, and has pads along only two sides. The dump drain allows for fast flushing, and the back bias allows the device to be made on 30\,$\mu$m thick epitaxial silicon or 100\,$\mu$m thick bulk silicon for higher red quantum efficiency (QE). 

The package is specifically designed for high resolution ultra stable spectroscopy. It has low thermal mass and directly integrates the temperature sensing and control with two different temperature control approaches.The preferred method has the servo control located beneath each of the two serial registers sandwiched between the flex circuit and the CCD package. A backup servo system is built into the backside of the CCD package.

The thermal constraints include a radiative load which is variable depending on surface properties and a conductive load that is linear with temperature through the thermal connections. A heat removal load is chosen to balance the cold sink in the center of the package backside at -100\,$^{\circ}$C. FEA analysis shows that the total load is expected to be 1.77\,W after settling and the temperatures in the package range from -98.1 to -99.5\,$^{\circ}$C. In contrast to other CCDs, the STA 4850 package is the only thermal mass. Transient thermal analysis shows that the system time constant is about 0.3 to 0.5 of a conventional CCD + cold block system, notably improving the temperature control of the detector.

\begin{figure}[!t]
\centering
\vspace{0mm}
\includegraphics[trim={0cm 3cm 0cm 3.5cm},width=0.49\linewidth,clip]{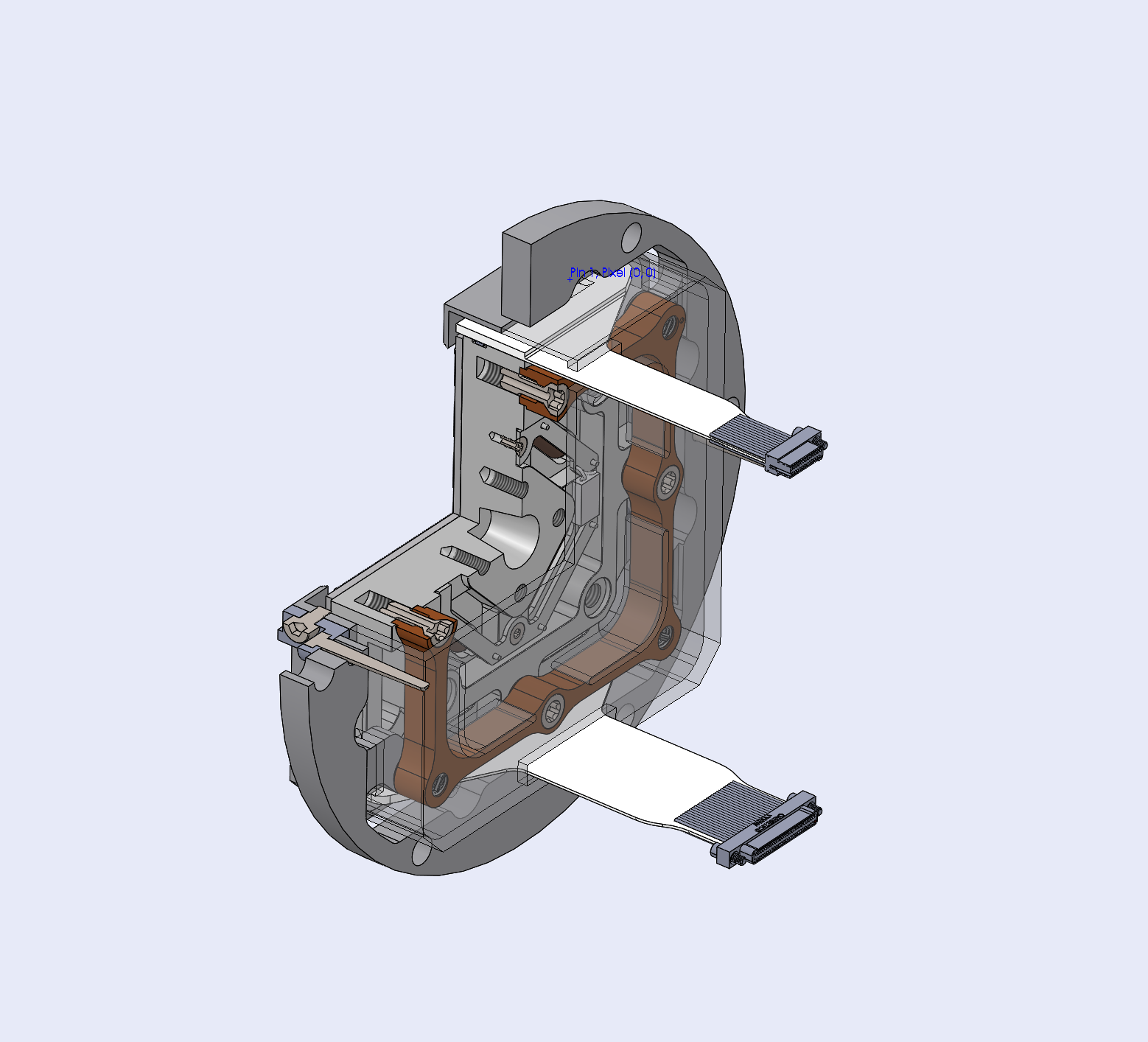}
\includegraphics[trim={0cm 0cm 0cm 0cm},width=0.49\linewidth,clip]{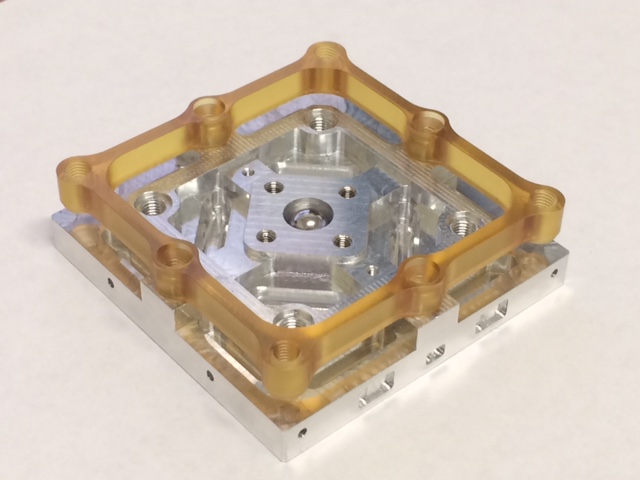}
\vspace{4mm}
\caption{\textbf{STA4850 CCD package}. \textsl{Left:} section view from the back side of the CCD package showing flex circuits (white), the back radiation shield (transparent), the thermal isolator (brown), the CCD package with back side servo, the Si CCD, the front radiation shield with integrated CCD mask, and the cryostat interface ring. The radiation shield was included to keep the CCD from sinking too much heat from the cryostat housing rather than the more typical use of creating a more efficient cooling system. The back side servo is a backup system in case the serial register servo under the flex circuits fails to work properly. The flex circuits include typical bias filtering circuits, and also include local JFET buffers which allow for higher speed operation with lower noise. For reference, the package is about 70 mm square, and the ring diameter is 116 mm.  \textsl{Right:} Prototype of the STA4850 package. The taper feature in the back serves as the connection for the cold sink. The taper multiplies the contact force and makes for the lowest possible thermal resistance with minimal tensile stress on the fasteners.}
\label{STA4850}
\end{figure}

The thermal isolator holds everything in place while also accounting for all of the temperature drop. It's the single part that does all of the stability work. Traditionally the isolator holds the CCD, cold plate, and other structural parts. With the 4850, the isolator is integrated into the package and the small constrained mass allows for a small isolator and little wasted heat. The total conductive heat load from the isolator is 151\,mW. The dominant conductive load is actually the copper in the flex circuit which is about 250\,mW.

The x/y stiffness (1/k) of the CCD package assembly is 0.16\,$\mu$m/N. In the less critical z (focus) dimension, the stiffness is 0.87\,$\mu$m/N. The CCD package weight is 0.95\,N. There are 3 modes of vibration that are essentially focus with the lowest frequency mode being pure focus at 223\,Hz. The next two modes are x/y translation and are at 621\,Hz. All modes are well above any excitation frequency from the cryocooler.
The CCD package also includes features allowing for a compact thermal isolator and radiation shield. These features minimize the thermal time constant and maximize mechanical stiffness. The mount itself is elastic and kinematic as well as being highly symmetric. See Figure \ref{STA4850} for a model and a prototype image of the CCD package.

We expect QE values of 90\% out to 900\,nm, whereas other, much thinner, deep-depletion devices commercially available in the desired detector format would drop to about 50\% at the same wavelength. The dump drain of the STA4850 device enables removal of charge from potential column defects before it hits the ADC. This feature becomes particularly important for thick deep-depletion devices with large wells used for long integrations. Large amounts of electrons from charge traps and hot columns can lead to detrimental effects at the ADC, affecting neighboring columns or even the whole image if the charge can not be dumped prior to the readout stage. 

\begin{SCfigure}[][!t]
\centering
\vspace{5mm}
\includegraphics[trim={2cm 4cm 0cm 0cm},width=0.6\linewidth,clip]{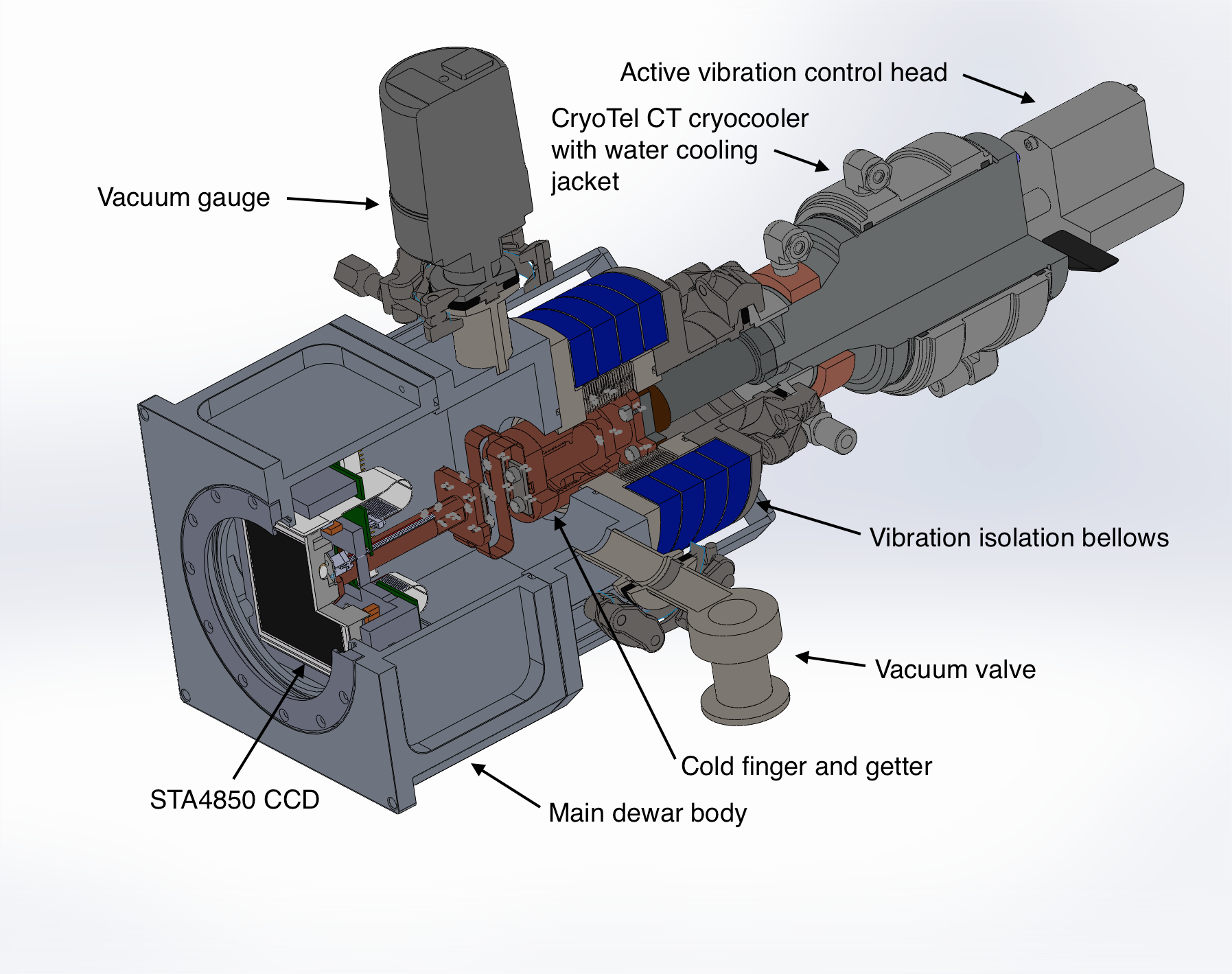}
\vspace{0mm}
\caption{\textbf{Cryostat model for the MAROON-X detector system.} The STA4850 CCD package is mounted in a small dewar body with flanges for a vacuum valve and a pressure gauge. At the backside of the dewar a CryoTel CT cryocooler is connected via a bellows feedthrough for vibration isolation. A getter and copper strap connect the cold tip of the cryocooler with the CCD package. An active vibration control head is mounted on the backside of the cryo\-cooler to actively measure and compensate the vibrations originating from the moving mass inside the Stirling-type cryo\-cooler.}
\label{cryostat}
\end{SCfigure}

The CCD cooling system will be based on a Stirling cryocooler, the CryoTel CT by SunPower. These compact coolers offer 11\,W of lift at 77\,K and a MTBF of over 200,000 hours. Excess heat is removed via chilled water and an external compressor with pressurized gas lines is no longer required. Strong vibrations, the only downside of these coolers, are mitigated via active vibration control (AVC) and additional bellows mechanically isolating the dewar from the cryocooler. See Figure \ref{cryostat} for details. 

\section{Fiber injection unit for the Gemini Telescope}\label{frontend}

The design of the fiber-injection unit (FIU) is driven by top-level requirements for throughput, image stabilization, and the ability to operate over a wide airmass range. The FIU optics have to provide high image quality and dispersion correction on the optical axis, but also sufficient performance in a moderate FOV for acquisition and to minimize stellar contamination of the sky background measurement, obtained by sky fibers approximately 20 arcsec off center. The FIU has to facilitate the conversion of Gemini's $f/16$ beam to $f/3.3$ to allow efficient fiber coupling with low FRD. The FIU has to provide access to the telescope pupil for an atmospheric dispersion corrector (ADC) and a tip-tilt (TT) mirror. It also has to facilitate re-imaging of the fiber focal plane for acquisition and guiding. Last but not least, the FIU needs to provide a way to align the chief rays of the telescope and the object fiber to within 10 arcmin to minimize angle-dependent FRD\cite{avila98}.

\begin{figure}[!t]
\centering
\vspace{5mm}
\includegraphics[trim={0cm 0cm 0cm 0cm},width=0.85\linewidth,clip]{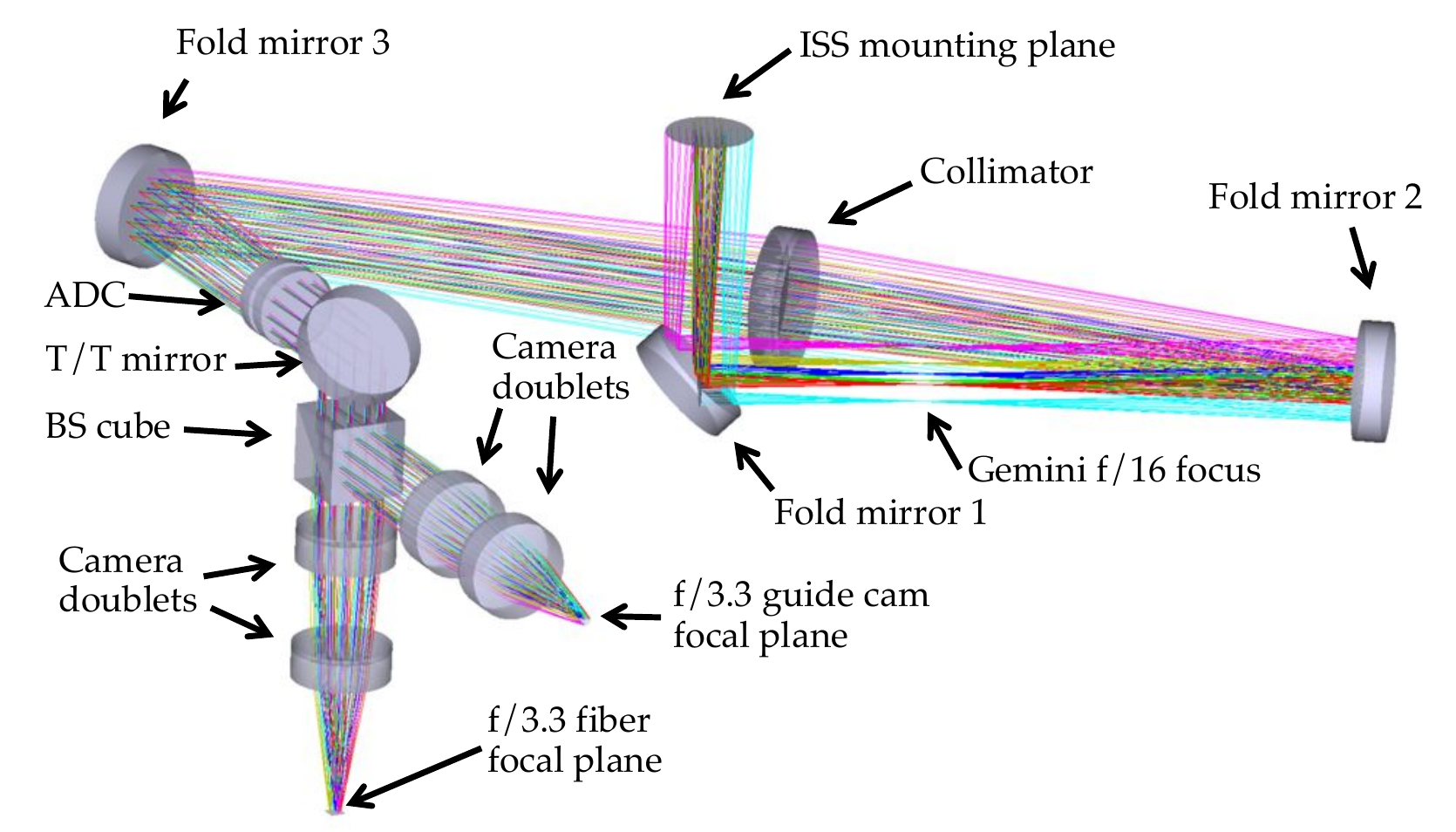}
\vspace{3mm}
\caption{\textbf{Optical layout of the fiber injection unit (FIU) at the Gemini Telescope}. The optical axis of the telescope is folded into a plane parallel to the primary mirror with the first fold mirror. The $f/16$ beam of Gemini is then collimated and passes through a classical ADC, which provides on-axis compensation of the atmospheric dispersion to under 50\,mas over 500--900\,nm down to air mass 2.5. A tip-tilt mirror in the pupil plane after the ADC facilitates differential guiding with a precision of $\le$0.1 arcsec. A beamsplitter cube after the TT mirror reflects about 1\% of the light towards a sCMOS camera for active guiding. The transmitted light is re-imaged onto a f/3.3 image plane, matching the acceptance cone of the 100\,$\mu$m octagonal object fiber. A retractable fold mirror between the ADC and fold mirror 3 (not shown here) is used to re-image the calibration fiber (embedded in the fiber plate) onto the science and sky fibers for calibration.}
\label{FIU}
\end{figure}

The FIU will be mounted on Gemini's Cassegrain port \#1, the direct (up-looking) port on the instrument support structure (ISS) behind the telescope's primary mirror. The FIU will use Gemini's peripheral wavefront sensor 2 (PWFS-2) for telescope guiding. Target acquisition and differential guiding corrections are done with the FIU's guide camera and TT mirror.

The optical design of the FIU is shown in Figure \ref{FIU} . The f/n system uses a collimator-camera combination in a near telecentric setup to convert the $f/16$ telescope beam to a $f/3.3$ beam at the fiber entrance. The collimator is a doublet made of S-FPM2 and S-BAM12. Each of the two cameras are comprised of two doublets made of S-FPL53, N-KZFS2 and S-BAM12. All surfaces are spherical with moderate ROCs. Diffraction limited performance is achieved on the optical axis and for a moderate FOV of 2 arcsec diameter. The f/n system provides a 1$\times$1 arcmin FOV which is just under 8$\times$8\,mm at $f/3.3$. It allows access to the telescope pupil for placement of a tip-tilt mirror and an ADC prism pair set in close proximity in front of the tip-tilt mirror (see also Figure \ref{FIUreal}).

The fibers at the $f$/3.3 focus are embedded in a fused silica plate. The 100\,$\mu$m octagonal object fiber is placed in the center of the nominal optical axis and two sky fibers are placed at a distance of 20 arcsec (2.6\,mm) to either side. Six additional fibers, three calibration fibers and three back-illuminated single-mode fibers, are placed around the object fiber. Fresnel losses on the fibers are minimized by bonding a BBAR coated flat on top of the fiber plate. All fold mirrors and the tip-tilt mirror have high-efficiency dielectric coatings and all refractive elements have a BBAR coating with an average reflectivity of R$\le$0.5\%. 

\begin{SCfigure}[][!t]
\centering
\vspace{5mm}
\includegraphics[trim={0cm 0cm 0cm 0cm},width=0.7\linewidth,clip]{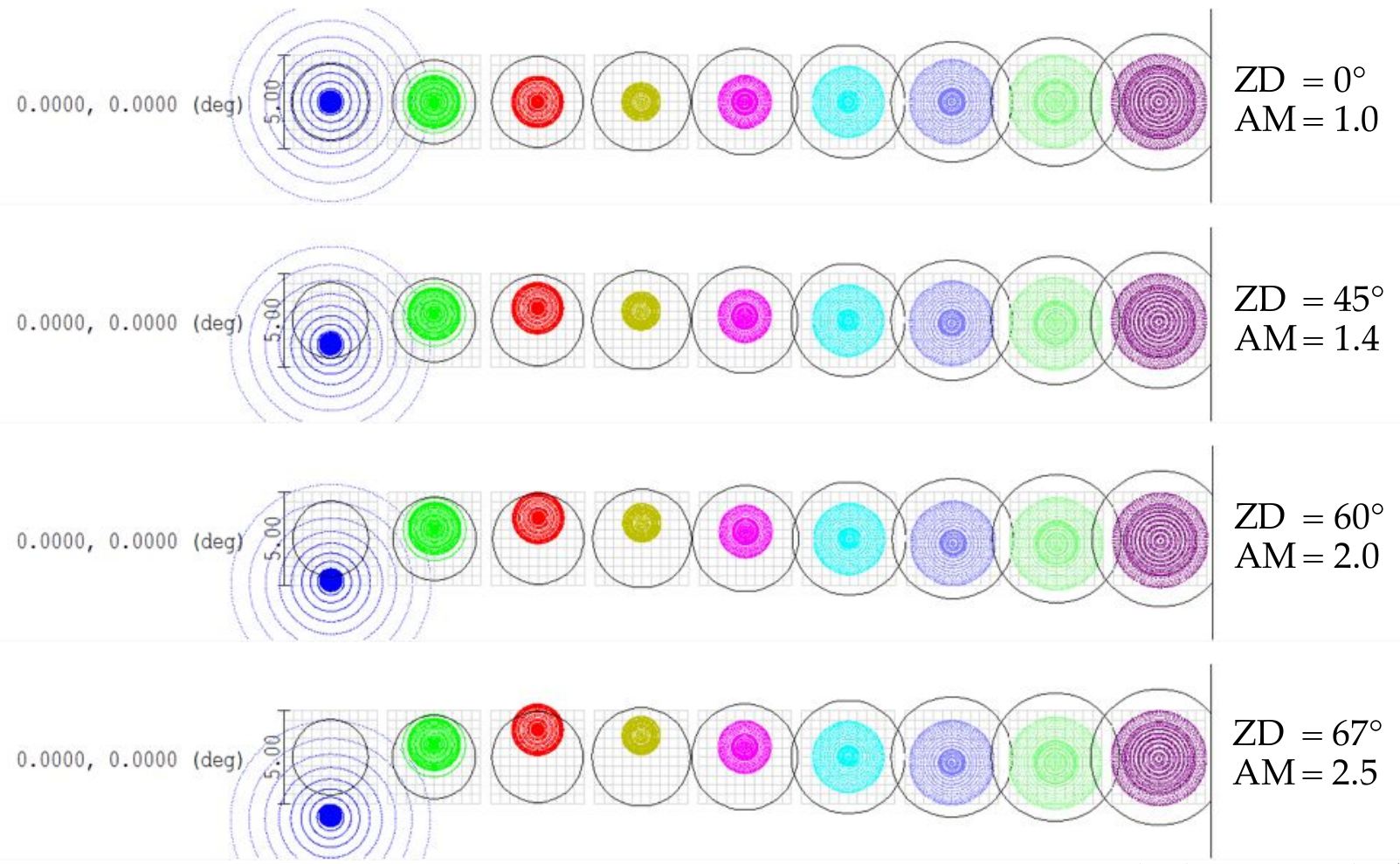}
\vspace{0mm}
\caption{\textbf{ADC performance for an on-axis target at various airmasses for 500 nm (left) to 900 nm (right)} in 50 nm increments. The box size is 5\,$\mu$m (40\,mas on sky), the circle shows the size of the Airy disc.}
\label{ADC}
\vspace{5mm}
\end{SCfigure}

\begin{figure}[!b]
\centering
\vspace{0mm}
\includegraphics[trim={0cm 0cm 0cm 0cm},width=0.9\linewidth,clip]{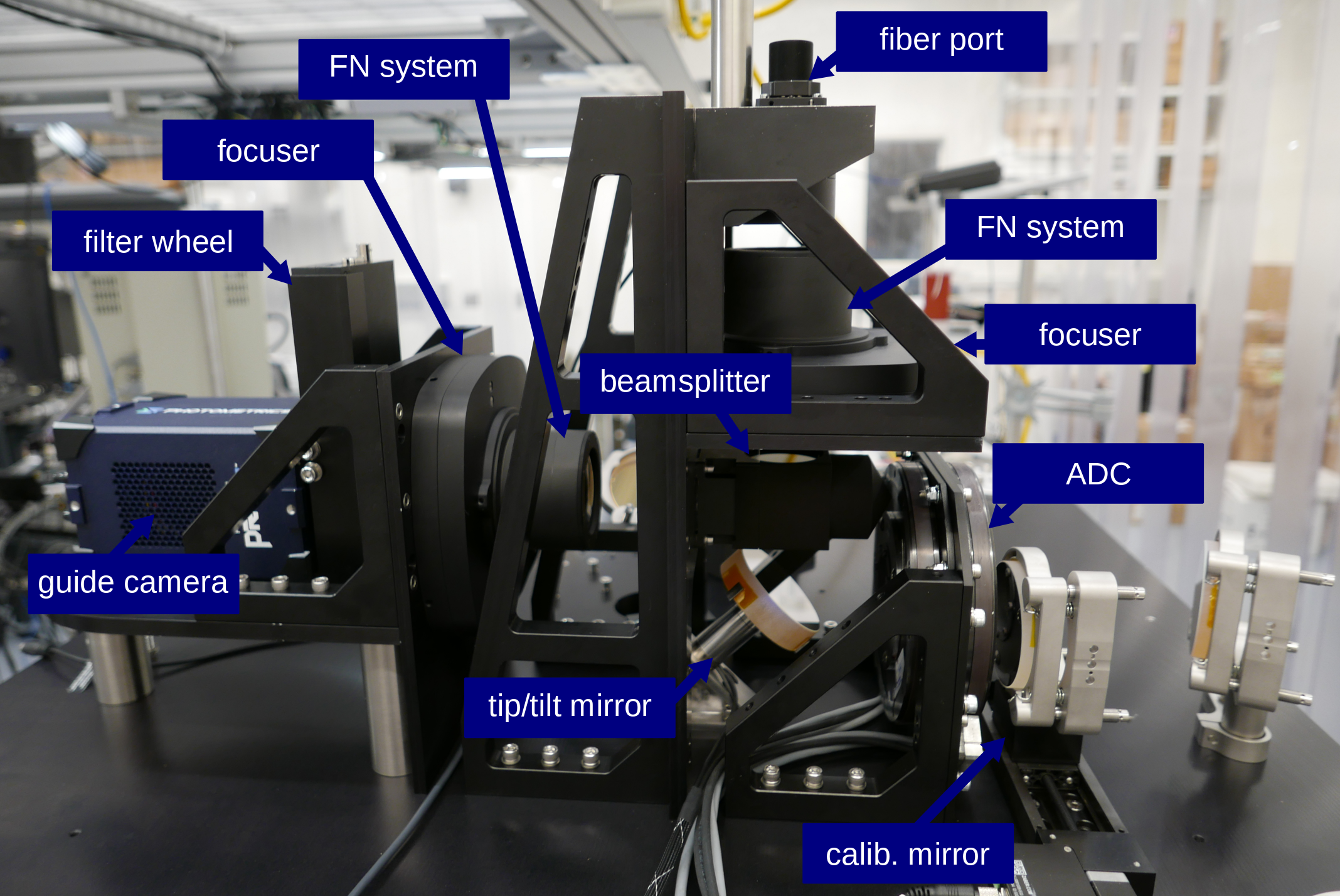}
\vspace{4mm}
\caption{\textbf{Image of the partially assembled fiber injection unit (FIU) for the Gemini Telescope}. The first two fold mirrors and the collimator are not visible in this image. The enclosure of the FIU has been completely removed.}
\label{FIUreal}
\end{figure}

The ADC is of a classical design with two identical counter-rotating Risley prism pairs made of two carefully selected glasses. The final design has a superb correction performance of $\le$50\,mas over 500--900\,nm down to air mass 2.5, very forgiving tolerances for prism separation, and a low chief ray angle steering of under 7 arcmin (see Figure \ref{ADC}). The off-axis performance out to several 10 arcsec is still within our original requirements of $\le$100\, mas for the maximum deviation over the full bandpass.

For the tip-tilt mirror we chose a stage from nPoint (model RXY6-210) with a closed-loop stroke of $\pm$3 mrad (0.17$^{\circ}$), and hence a throw of $\pm$6.3 arcsec ($\pm$4.4 arcsec) on sky in X and Y, respectively. It is thus desirable to offload offsets larger than about 2 arcsec to the telescope, particularly during object acquisition. Due to the weight of the mirror, the update frequency of the tip-tilt stage is limited to a few Hz. This is still fast enough to correct for flexure inside the FIU, compensate for thermal drifts and residual beam steering of the ADC, and to steer the image of the calibration fiber to either the object or the sky fiber during calibration exposures.

\vspace{5mm}

A beam splitter comprised of two rectangular prisms bonded with a UV adhesive that has a refractive index mismatch of about $\Delta n=0.13$ reflects about 1--2\% of the light towards a sCMOS guide camera. We are thus not guiding on the extended wings of the PSF but on the core of the PSF, directly imaged by the guide camera. We have also bonded a corner-cube prism to the face of the beamsplitter cube opposite of the guide camera. The corner cube re-directs about 1\% of the light exiting from the three single mode fibers surrounding the object fiber in the fiber plate towards the guide camera. These three points can be used to precisely triangulate the actual position of the object fiber in real time. By illuminating the single-mode fibers with light outside the spectrograph bandpass, we avoid potential contamination of the science and sky spectrum. 

We choose a sCMOS detector system from Photometrix for our guide camera. The detector is a GSENSE 2020 BSI with 2k$\times$2k 6.5\,$\mu$m pixels. The QE peaks at 95\% and the median read noise is 1.3\,e$^-$. Full frame read out speeds of over 40\,fps at 16\,bit can be achieved.  

All refractive optical elements, except for the beam splitter prisms and the retro-reflector were manufactured by Optimax. The collimator, the cameras, the ADC prisms, and the beamsplitter cubes are mounted in custom SS\,416 mounts, which offered the best compromise in terms of CTE match to the various glasses. We use COTS mounts for the fold mirrors. The camera barrels are mounted in two commercial focuser units that allow sub-micron precision control of the camera focus position. 

Since the FIU is mounted on the Gemini Cassegrain port, it will be subject to varying gravity vector orientations. We have thus designed the FIU opto-mechanics to be both lightweight and stiff. Remaining flexure of the FIU during telescope and de-rotator movements will be compensated by the active guiding system and the tip-tilt mirror. MAROON-X will initially be a visitor instrument operated in campaign mode and the FIU will regularly be swapped against another Gemini instrument. To support these swaps, the fiber feed is easily detachable from the FIU. Locator pins on the FIU permit removal and re-installation on Gemini's instrument support structure (ISS) without the need for re-alignment of the instrument.

A direct fiber run from the FIU through the central hole in the telescope pier down in the pier lab where MAROON-X will be installed, allows for a short fiber run of under 30\,m. 
\begin{SCfigure}[][!b]
\centering
\vspace{0mm}
\includegraphics[trim={0cm 0cm 0cm 0cm},width=0.6\linewidth,clip]{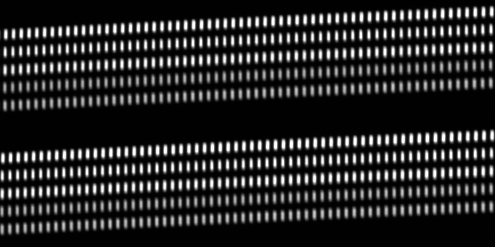}
\vspace{0mm}
\caption{\textbf{Small section of a calibration frame} showing all five rectangular fibers of two echelle orders illuminated with our Rb-traced white light etalon. The etalon has a FSR of 15\,GHz, which at this wavelength equals approximately 8 pixels. The lines are unresolved, thanks to an etalon finesse of about 30. The difference in brightness between the individual fibers is an illumination artefact at the calibration unit and will be removed when the pupil slicer is installed. }
\label{etalonframe}
\end{SCfigure} 

\section{First RV stability tests}

After the installation of the core spectrograph in the environmental chamber in the lab at the University of Chicago, we integrated the instrument with our calibration unit which includes a Rb-traced white light etalon\cite{etalon}. Since the red arm and the science detector systems are not yet delivered, we installed an off-the-shelf interim detector system on the blue arm. We use a FLI ProLine PL230 with a 2k$\times$2k 15\,$\mu$m e2v CCD 230-42 and replaced the dewar window with a smaller version of our field-flattener lens, using a modified window holder. This setup was used for initial acceptance testing and is in use for ongoing RV stability tests until the final science grade detector systems arrive.

By bypassing the pupil slicer, we have access to all five rectangular fibers individually and can test their RV performance separately to determine the best choice for the calibration and sky fiber relative to the three object fibers (see Figure \ref{etalonframe}).

RV stability measurements at the sub-m\,s$^{-1}$ level require a sophisticated software package. A first version of this package is operational and there are ongoing efforts to optimize it for eventual delivery with the instrument. Initial RV tests revealed a larger than expected instrumental response to external temperature variation inside the environmental chamber with a clear correlation between absolute RV zero point and the temperature of the camera and detector mount. Efforts are underway to isolate the responsible part and fix the issue. 

\begin{SCfigure}[][!t]
\centering
\vspace{0mm}
\includegraphics[trim={0cm 0cm 2cm 1cm},width=0.6\linewidth,clip]{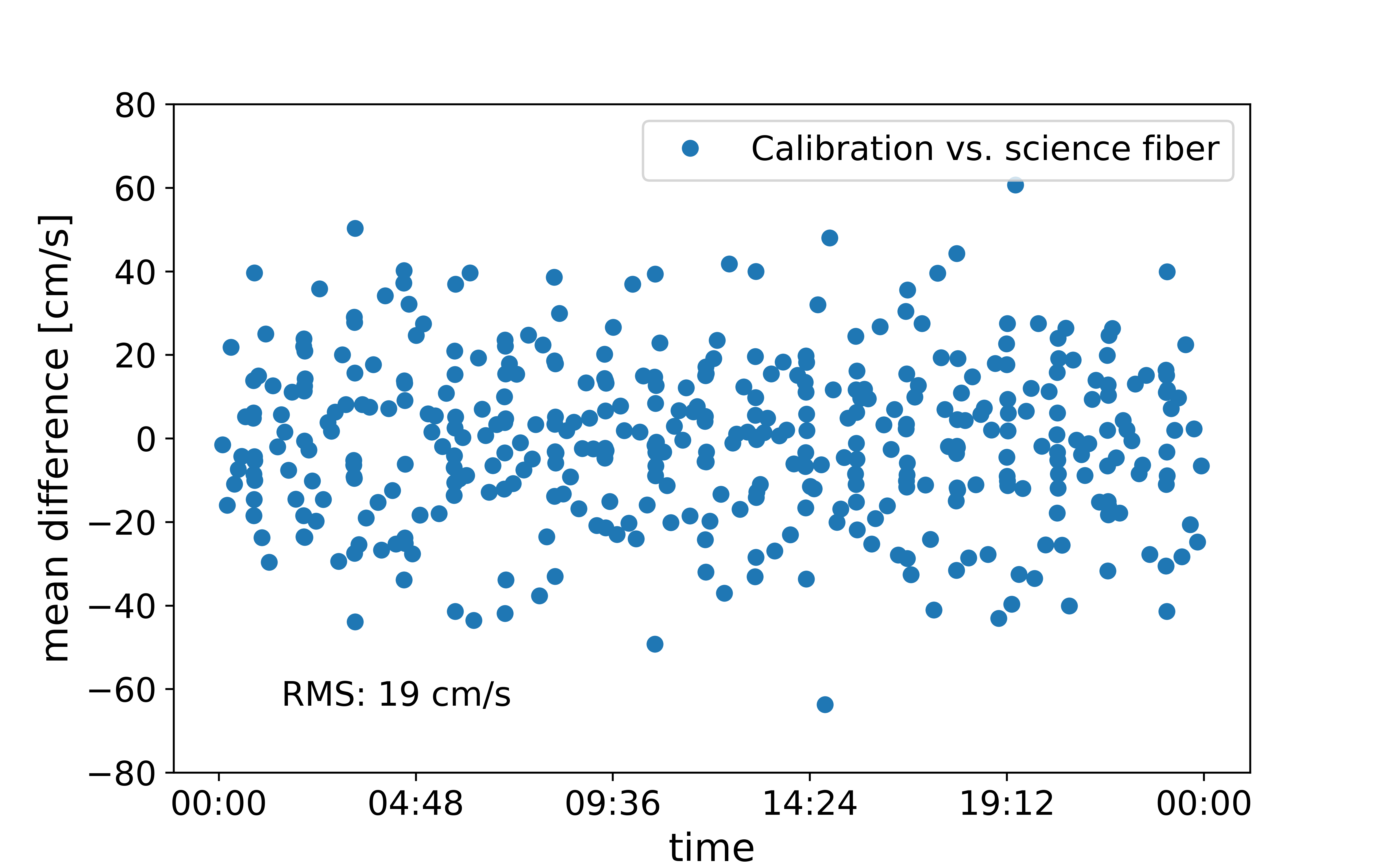}
\vspace{0mm}
\caption{\textbf{Difference in radial velocity between one of the three science fibers and the calibration
fiber} for approximately 4,000 etalon lines from 16 orders of the blue arm on the interim 2k$\times$2k e2V detector. The timespan is 24 hours. Data were taken with 10 second to 5 minute cadence. No drift between
the two fibers is apparent at the cm/s level with an RMS of 19 cm/s for the individual exposures.}
\label{RV}
\end{SCfigure}

\begin{table}[!b]
\caption{MAROON-X main characteristics}
\begin{center}
\begin{tabular}{|l|l|}
\hline
Spectral resolution & R = 80,000 \\
Acceptance angle & FOV = 0.77'' at the 8\,m Gemini North Telescope \\
Wavelength range & 500 nm -- 900 nm (in 56 orders)\\
Number and reach of arms & 2 (500--670\,nm and 650--900\,nm) \\
Cross-disperser & anamorphic VPH grisms\\
Beam diameter & 100\,mm (at echelle grating), 33\,mm (at cross-disperser)\\
Main fiber & 100\,$\mu$m octagonal (CeramOptec)\\
Number and type of slicer & 3x pupil slicer \\ 
Slit forming fibers & 5$\times$ 50$\times$150\,$\mu$m rectangular (CeramOptec)\\
Inter-order and inter-slice spacing & $\geq10$ pixel \\
Average sampling & 3.5 pixel per FWHM\\
Blue detector & Standard epi 30\,$\mu$m thick 4k$\times$4k STA4850 CCD (15\,$\mu$m pixel size)\\
Red detector & Deep-depletion 100\,$\mu$m thick 4k$\times$4k STA4850 CCD (15\,$\mu$m pixel size)\\
Calibration & Fabry-P\'erot etalon for simultaneous reference (fed by 2nd fiber) \\
Environment for main optics & Vacuum operation, 1\,mK temperature stability\\
Environment for camera optics & Pressure sealed operation, 20\,mK temperature stability\\
Long-term instrument stability & \SI{0.7}{\mps} (requirement), \SI{0.5}{\mps} (goal)\\
Total efficiency & 11\% (requirement) to 15\% (goal) at \SI{700}{\nano\meter} (at median seeing)\\
Observational efficiency & S/N=100 at \SI{750}{\nano\meter} for a V=16.5 late M dwarf in \SI{30}{\minute} \\
\hline
\end{tabular}
\end{center}
\label{table}
\vspace{0mm}
\end{table}

However, inter-fiber drift, i.e.\ relative RV measurements between the science and the calibration fiber, mimicking the actual RV observations during science observations, is not observed down to the cm\,s$^{-1}$ level over a 24\,hr timespan. The rms of the individual RV measurements, taken at 10\,s to 5\,min cadence is only 19\,cm\,s$^{-1}$, see Figure \ref{RV}. This is an encouraging result, given the limitations of this test. The spectral coverage of the 2k$\times$2k interim detector is much smaller than the final science grade detectors. Its  temperature stability is severely limited by its Peltier cooling system and the spectrograph was still settling towards its temperature setpoint. Last but not least, the pupil slicer and double scrambler was not used and we relied only on the scrambling power of the octagonal and rectangular fibers feeding the spectrograph. We are thus confident that the instrument-limited end-to-end RV stability goal of 0.5\,m\,s$^{-1}$ can be achieved.

\section{Outlook}

After the integration of the red arm and the science grade detector systems in the fall of 2018, RV stability tests and software development will continue until the system is ready for deployment. We estimate that MAROON-X will be installed at Gemini North in the first quarter of 2019. The fiber injection unit and the environmental chamber will already be installed and commissioned in late 2018 so that these systems are functioning and ready before the spectrograph is installed. First science observations are foreseen for the second half of semester 2019A.

\acknowledgments     
 
The University of Chicago group acknowledges funding for this project from the David and Lucile Packard Foundation through a fellowship to J.L.B., as well as support from the Heising-Simons Foundation, and from the University of Chicago.



\bibliographystyle{spiebib}   

\end{document}